\documentclass[dvips]{article}
\usepackage{icrctc07}

\title{Measurement of the angular resolution of the ARGO-YBJ detector}
\shorttitle{Measurement of the angular resolution}
\authors{G. Di Sciascio$^{1}$ and E. Rossi$^{1}$ for the ARGO-YBJ Collaboration}
\shortauthors{G. Di Sciascio$^{1}$ et al.}

\afiliations{$^{1}$INFN, sez. di Napoli, Italy }
\email{elvira.rossi@na.infn.it; giuseppe.disciascio@na.infn.it}

\abstract{The ARGO-YBJ experiment is a full coverage EAS-array
installed at the YangBaJing Cosmic Ray Laboratory (4300 m a.s.l.,
Tibet, P.R. China). We present the results on the angular
resolution measured with different methods with the full central
carpet. The comparison of experimental results with MC simulations
is discussed.}
\begin{document}
\maketitle
\section{Introduction}
The ARGO-YBJ detector is constituted by a single layer of
Resistive Plate Chambers (RPCs). This carpet has a modular
structure, the basic unit is a cluster, composed by 12 RPCs
(2.8$\times$1.25 m$^2$ each). Each chamber is read by 80 strips,
logically organized in 10 independent pads\cite{nim_argo}. The
central carpet, constituted by 10$\times$13 clusters with
$\sim$93$\%$ of active area, is enclosed by a guard-ring partially
instrumented ($\sim$40$\%$) in order to improve rejection
capability for external events. A lead converter 0.5 cm thick will
uniformly cover the apparatus in order to improve the angular
resolution. Since December 2004 the pointing accuracy of the
detector has been studied, during the detector setting-up, with 3
different carpet areas: 42 clusters (ARGO-42, $\sim$1900 m$^2$),
104 clusters (ARGO-104, $\sim$4600 m$^2$) and the full central
carpet, 130 clusters (ARGO-130, $\sim$5800 m$^2$), yet without any
converter sheet. The data have been collected with a so-called
{\it "Low Multiplicity Trigger"}, requiring at least $20$ fired
pads on the whole detector.

\section{Estimate of the angular resolution}
Searching for cosmic $\gamma$-ray point sources with ground-based
arrays the main problem is the rejection of the background of
charged cosmic rays, therefore a good angular accuracy in
estimating the arrival direction is necessary. The angular
resolution has in general two components: a statistical one, due
to fluctuations of the shower development and the detector noise,
and a systematic error (i.e., the pointing error) arising from a
possible misalignment of the detector, an asymmetry of the array
geometry and some systematic bias induced in the shower
reconstruction process (systematic error on core position
determination, the change of EAS front conical slope with size,
etc.). The standard method to estimate the statistical angular
resolution of an EAS array is the so-called "Chessboard Method".
The pointing error, instead, can be studied by observing the
shadowing effect of cosmic rays from the Moon direction. Other
systematic errors can be investigated by means of MC simulations
comparing the true and reconstructed primary directions. In this
paper we report on the angular resolution of the increasing
ARGO-YBJ detector with the following techniques: (1) chessboard
method, which splits the detector into two parts and compares the
two measured arrival directions; (2) MC simulation; (3)
a preliminary study of the Moon shadow.\\

{\bf Event reconstruction}\\
%
%%%%%%%%%%%%%%%%%%%%%%%%%%%%%%%%%%%%%%%%%%%%%%%%%%%%%%%%%%%%%%%%%%%%
%%%%%%%%%%%%%%%%%%%%%%%%%%%%%%%%%%%%%%%%%%%%%%%%%%%%%%%%%%%%%%%%%
\begin{figure}
\begin{minipage}[t]{1\linewidth}
  \begin{center}
\includegraphics[width=0.85\textwidth,angle=0,clip]{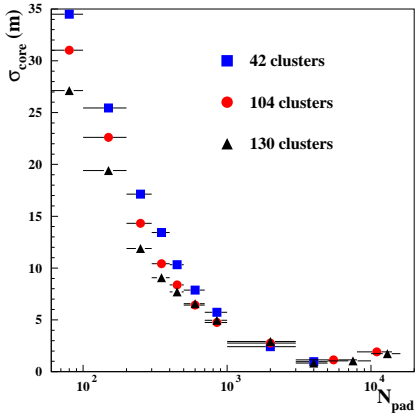}
\caption{\label {coreres} Shower core position resolution of
internal selected protons as a function the pad multiplicity for different detector dimensions.}
  \end{center}
 \end{minipage}\hfill
\end{figure}
To find the optimal selection method we have to rely on MC
calculations, thus we have simulated, via the Corsika/QGSJet code
\cite{corsika}, proton-induced showers with particle spectrum
$\propto E^{-2.78}$ ranging from 300 GeV to 1 PeV and a Crab-like
spectrum $\propto E^{-2.49}$ for photons ranging from 300 GeV to
100 TeV. The detector response has been simulated via a
GEANT3-based code. The core positions have been randomly sampled
in an energy-dependent area large up to $1000 \times 1000~m^2$,
centred on the detector. For ARGO-130, showers are considered
internal if they satisfy the following condition: the particle
density in the inner $8 \times 11$ clusters is higher than that of
the outer ring constituted by 42 clusters. The shower core
positions of the selected events are hence reconstructed by means
of the Maximum Likelihood Method: any core lying outside the
detector edge is further rejected. In Fig.\ref{coreres} the shower
core position resolution of internal selected protons is shown for
ARGO-42, ARGO-104 and ARGO-130 detectors. The resolution worsens
due to the detector saturation at very large shower sizes (the
total pad number goes from 5040 for ARGO-42 to 15600 for
ARGO-130). For details about the analysis with smaller carpets see
\cite{lisbona}. From the figure it results that the core position
is reconstructed with a resolution better than $2~m$ for
$N_{pad}\geq 1000$ (median energy $E_{p}\sim 10~TeV$). The
majority of the incorrectly accepted and rejected events are
located near the carpets boundary, making the contamination less a
concern, as the core of these events can still be located with
small errors.

{\bf Analysis with the Chessboard Method}\\
In this analysis the shower primary direction is reconstructed by
means of an iterative procedure, with a conical correction to the
shower front fixed to the value $\alpha$ = 0.03 ns/m, applied to
events reconstructed inside the carpet area. The relative time
offset (due to differences in cable length, etc.) among different
pads has been estimated with the so called ``Characteristic Plane
Method''\cite{cal_1,cal_2}. The analysis presented in this paper
refers to showers with a zenith angle $\theta<$15$^{\circ}$. About
$\sim 10^7$ events have been selected and analyzed with the
procedure described in the previous section. We require that the
difference in the number of fired pads in both sub-arrays must be
less than 10$\%$. This guarantees that both reconstructions have a
similar systematical and statistical error. In order to estimate
the pointing accuracy of the detector we used the $\psi_{72}$
parameter, a measure of the angular resolution defined as the
value in the angular distribution which contains $\sim$72$\%$ of
the events. This is a useful definition because, assuming that the
Point Spread Function (PSF) for the entire detector is a Gaussian,
it describes a solid angle which maximizes the signal/background
ratio from a point source on a uniform background
\cite{protheroe}. The rms projected angular resolution of the
detector is given by the relation $\sigma_{\theta} \approx
\psi_{72}/1.58$.
\begin{figure}
\begin{minipage}[t]{1\linewidth}
  \begin{center}
\includegraphics[width=0.85\textwidth,angle=0,clip]{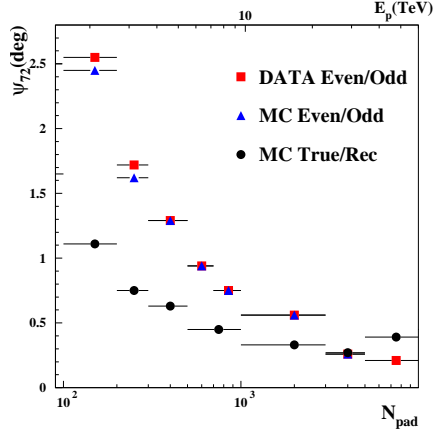}
\caption{\label{evenodd_mcdata} The opening angle $\psi_{72}$ as a
function of pad multiplicity measured with ARGO-130 compared with
MC simulations. The upper scale shows the estimated median energy
for proton-induced showers. The error bars refer to the width of
the pad multiplicity bins. }
  \end{center}
 \end{minipage}\hfill
\end{figure}
In Fig.\ref{evenodd_mcdata} the opening angle $\psi_{72}$ for
ARGO-130 calculated via the chessboard method with data is
compared, as a function of pad multiplicity $N_{pad}$ (i.e., the
sum of even and odd pads), to the MC simulation. As it can be seen
from the plot, there is a satisfactory agreement of the simulated
result with the experimental one. The $\psi_{72}$ parameter
improves roughly proportionally to $N_{pad}^{-0.7}$ for ARGO-42,
ARGO-104 \cite{lisbona} and ARGO-130. In a shower flat temporal
profile approximation, neglecting any dependence on the core
position, one would expect, on a simple statistical basis, that
$\psi_{72}$ decreases as $N_{pad}^{-0.5}$. However, as the
increased number of fired pads also means an increased shower
size, and therefore an increased number of particles detected on
the single pad, the intrinsic error in timing (due to the disc
thickness and curvature) decreases, leading to a steeper than
$N_{pad}^{-0.5}$ behaviour in the overall angle
estimate.\\

\begin{figure}
\begin{minipage}[t]{1\linewidth}
\begin{center}
\includegraphics[width=0.85\textwidth,angle=0,clip]{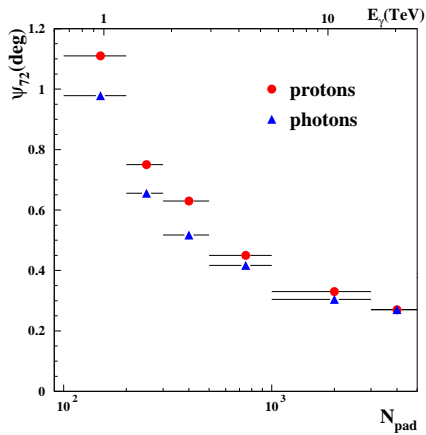}
\caption{\label{psi70_fotoni}The opening angle $\psi_{72}$ as a
function of pad multiplicity for protons and photons. The error
bars refer to the width of the pad multiplicity bins. The upper
scale shows the estimated median energy for photon-induced
events.}
\end{center}
\end{minipage}\hfill
\end{figure}

{\bf Analysis with the MC simulation}\\
The true shower direction of the MC events is known, therefore the
angular resolution can be computed directly from the differences
$\Delta\theta_{true/rec}$ between true and reconstructed shower
directions. In Fig.\ref{evenodd_mcdata} the filled circles refer
to the parameter $\psi_{72}$ calculated via MC simulations. The
opening angle worsens due to the detector overflows at very large
shower sizes (a behaviour similar to that of the shower core
position resolution in Fig. \ref{coreres}). Unlike the chessboard
method, the calculation of the angular resolution in this case is
sensitive to the shower core position resolution and to the
accuracy of the temporal profile description. As a consequence,
these systematic errors can be limiting factors for
$\Delta\theta_{true/rec}$. If the two sub-arrays are totally
independent, the even-odd angular difference is expected to be
approximately twice the angular resolution of the entire detector:
$[\sigma_{true/rec}]/[\sigma_{eo}]\sim 0.5$ \cite{alexandreas}. As
it can be seen from Fig.\ref{evenodd_mcdata}, this hypothesis is
not correct: a dependence of the ratio
$[(\psi_{72})_{true/rec}]/[(\psi_{72})_{eo}]$ on the total pad
multiplicity is evident. This ratio varies from $\sim$0.5 for very
small showers to $\sim$1 for large showers (N$_{pad}\sim$ few
thousands). This is due to the effect of systematical errors which
add quadratically to the statistical ones estimated by the
chessboard method. At very low multiplicity the effect of the
statistical errors is dominant and
$[(\psi_{72})_{true/rec}]/[(\psi_{72})_{eo}]\sim$ 0.5. When this
ratio is about 0.7 the systematical and statistical errors are
equivalent. As a consequence, we have calculated with a simulation
the factor by which the measured angle $(\psi_{72})_{eo}$ must be
multiplied to obtain the angular resolution. As an example, the
{\em average statistical angular resolution} for the ARGO-130
detector measured with the chessboard method, up to $\approx$4000
fired pads ($E_p \sim 30~TeV$), can be described by the following
equation:\\
$$\sigma_{eo}(deg)=\frac{(\psi_{72})_{eo}}{1.58}\cdot [0.42+1.4\cdot10^{-4}~N_{pad}].$$
Obviously, this measured angular resolution refers to
proton-induced air showers. The angular resolution for
photon-induced showers is slightly lower due to their better
defined temporal profile at low multiplicities. The opening angle
$\psi_{72}$ as a function of pad multiplicity for protons and
photons is compared in Fig. \ref{psi70_fotoni}.

Another probable source of systematical error may be an inaccurate
shower profile description. Indeed, as it is well known, the
conical slope of the shower front lowers with increasing shower
size. These systematical errors affect both directions
reconstructed by the sub-arrays in the same way, tilting the
result in the same direction. In view of making conservative
estimates of the angular resolution for N$_{pad}\geq$ 4000
we use the worse resolution, i.e. that determined via MC
simulations: $\sigma\approx$ 0.2$^{\circ}$.
The addition of a 0.5 cm lead sheet on
top of the RPCs will lead to an improvement of the angular
resolution by a factor of at least 30$\%$ for low pad
multiplicity (below some hundreds fired pads)\cite{bacci}.\\

{\bf Analysis with the shadow of the Moon}\\
The analysis of the deficit of cosmic rays from the direction of
the Moon is a well known method to determine the angular
resolution and the systematical pointing error of an EAS array
based on the deficit profile and the peak shift from the Moon
position.
%
%%%%%%%%%%%%%%%%%%%%%%%%%%%%%%%%%%%%%%%%%%%%%%%%%%%%%%%%%%%%%%%%%
\begin{figure}
\begin{minipage}[t]{1\linewidth}
\begin{center}
\includegraphics[width=0.71\textwidth,angle=0,clip]{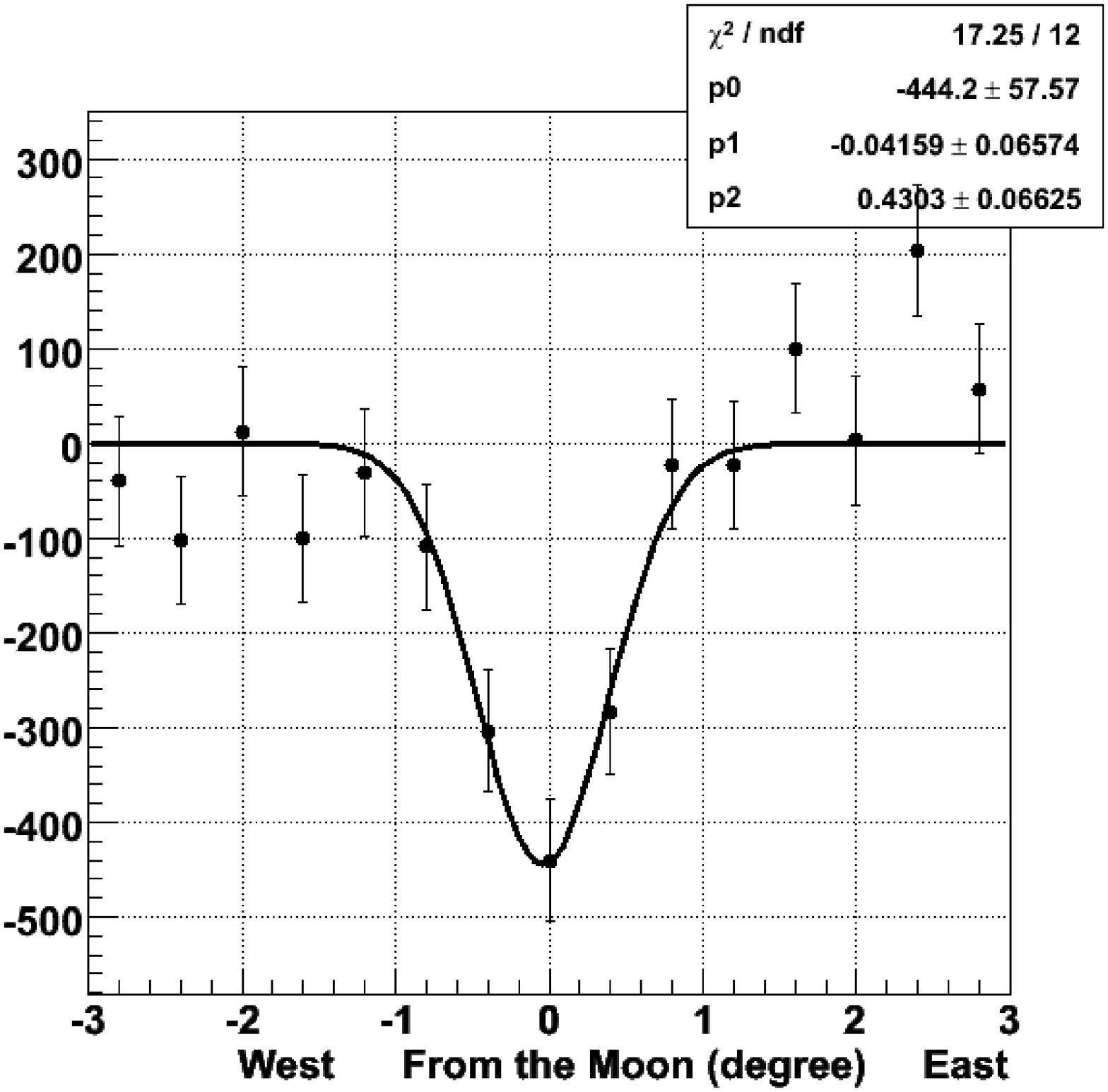}
\end{center}
\end{minipage}\hfill
\begin{minipage}[t]{1\linewidth}
\begin{center}
\includegraphics[width=0.71\textwidth,angle=0,clip]{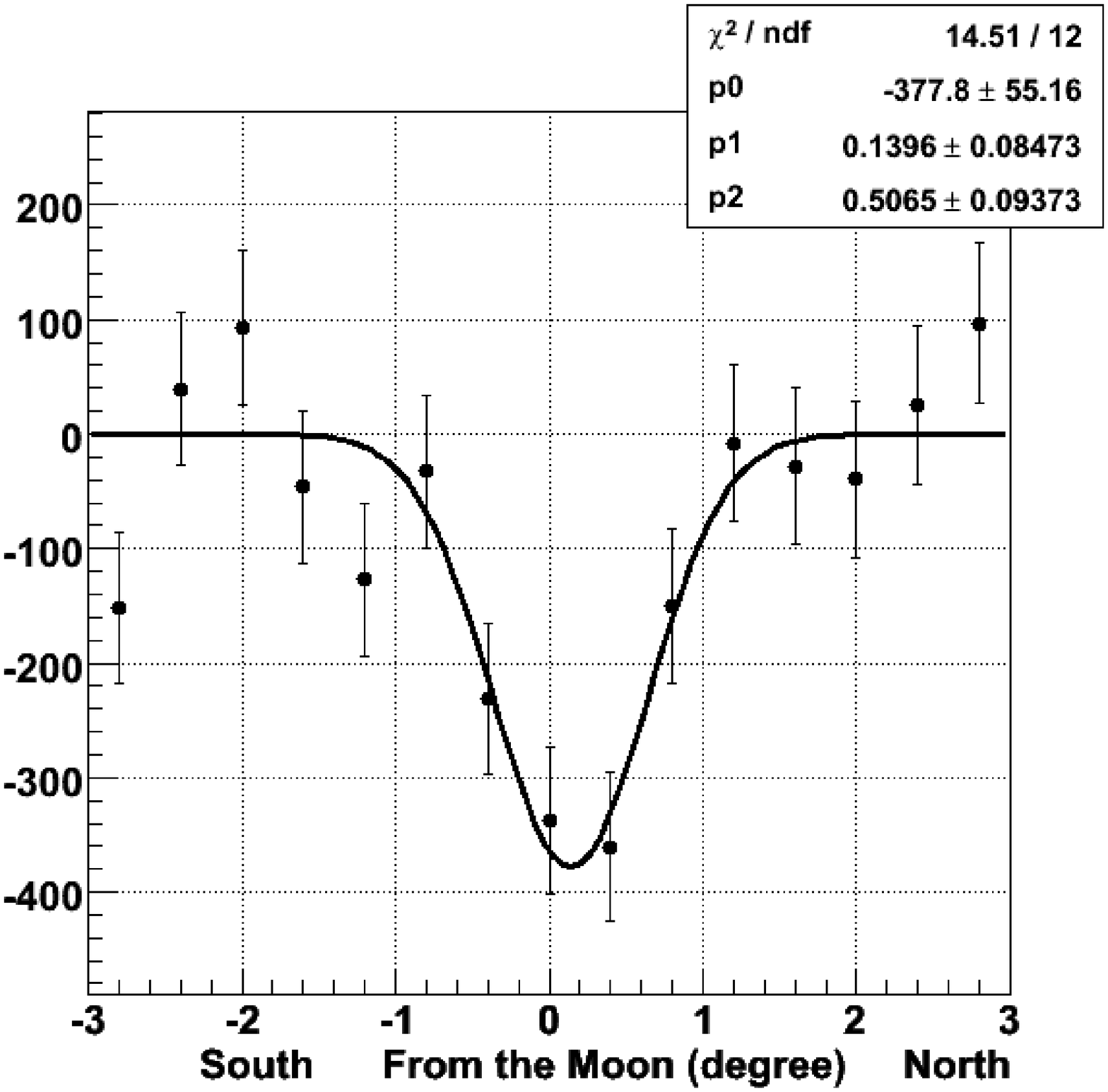}
\end{center}
\end{minipage}\hfill
\caption{\label{42cl_Moonskymap} The distribution of observed
deficit event number projected to the W-E and N-S axes for
N$_{pad}>$500.}
\end{figure}
%%%%%%%%%%%%%%%%%%%%%%%%%%%%%%%%%%%%%%%%%%%%%%%%%%%%%%%%%%%%%%%%%%%%
%
From July 2006 to February 2007 ARGO-130 observed the Moon for
$\sim$558 h. A very preliminary analysis of the shadow of the Moon
has been performed filling a 2-dimensional sky map around the Moon
position \cite{icrc07_skymap}. The statistical significance of the
deficit of cosmic ray events is $\approx$ 11$\sigma$ for
N$_{pad}>$120 (E$_{p}\approx$ 3 TeV). We note that the low energy
threshold and the pointing accuracy of the detector lead to a Moon
shadow detection in a very short observation time. As an example,
in Fig. \ref{42cl_Moonskymap} the distribution of observed deficit
event number projected to the W-E and N-S axes are shown. The
events are selected with a pad multiplicity $> 500$
(E$_{p}\approx$ 5 TeV) and with the core reconstructed inside the
ARGO-130 boundaries. The projections are fitted with a Gaussian
function with $\sigma_{W-E}=0.43^{\circ}$,
$\sigma_{N-S}=0.51^{\circ}$, consistent with MC calculations. A
residual systematic pointing error of about $0.14^{\circ}$ is
visible in N-S direction.
\section{Conclusions}
Since December 2004 increasing fractions of ARGO-YBJ detector
have been put in data taking even with a reduced duty-cycle due
to installation and debugging operations.
In this paper we
presented a measurement of the pointing accuracy of the ARGO-130 detector.
The capability of reconstructing the
primary shower direction has been investigated with the chessboard
method and with a preliminary Moon shadow analysis.
Studies are in progress in order to determine the final angular resolution.
%
%
%
%This is the reference to .bib file (Whitout .bib!)
%\bibliography{libros}
%This in the bibtex style, is ok.
%\bibliographystyle{plain}
%
%

\end{document}